\documentclass[aps,prd,amsmath,amssymb,superscriptaddress,preprintnumbers,twocolumn,10pt,nofootinbib]{revtex4-1}
\usepackage{graphicx}
\usepackage{dcolumn}
\usepackage{bm}
\usepackage{amssymb}
\usepackage{latexsym}
\usepackage{booktabs}
\usepackage{amsmath}
\usepackage{multirow}
\usepackage{url}
\usepackage{footnote}
\usepackage{float}
\usepackage{threeparttable}
\usepackage[bottom]{footmisc}

\usepackage[normalem]{ulem}
\usepackage{color}
\usepackage{array}
\usepackage{enumerate}
\usepackage{adjustbox}
\usepackage{makecell}
\usepackage{diagbox}
\usepackage{epstopdf}
\usepackage{epsfig}
\usepackage{longtable}
\usepackage{supertabular}
\usepackage{algorithm}
\usepackage{pifont}
\usepackage{algorithmic}
\usepackage{changepage}
\usepackage{setspace}
\usepackage[colorlinks=true, linkcolor=blue, citecolor=blue]{hyperref}

\begin{document}

\title{Alleviating the $H_{0}$ tension through the interacting dark energy model from quantum gravitational field theory in light of DESI DR2}

\author{Yi-Min Zhang} 
\affiliation{Liaoning Key Laboratory of Cosmology and Astrophysics, College of Sciences, Northeastern University, Shenyang 110819, China}

\author{Tian-Nuo Li}  
\affiliation{Liaoning Key Laboratory of Cosmology and Astrophysics, College of Sciences, Northeastern University, Shenyang 110819, China}

\author{Guo-Hong Du} 
\affiliation{Liaoning Key Laboratory of Cosmology and Astrophysics, College of Sciences, Northeastern University, Shenyang 110819, China}

\author{Sheng-Han Zhou} 
\affiliation{Liaoning Key Laboratory of Cosmology and Astrophysics, College of Sciences, Northeastern University, Shenyang 110819, China}

\author{Li-Yang Gao} 
\affiliation{School of Science, Jiangxi University of Science and Technology, Ganzhou 341000, China}

\author{\\Jing-Fei Zhang} 
\affiliation{Liaoning Key Laboratory of Cosmology and Astrophysics, College of Sciences, Northeastern University, Shenyang 110819, China}

\author{Xin Zhang}\thanks{Corresponding author}\email{zhangxin@mail.neu.edu.cn}   
\affiliation{Liaoning Key Laboratory of Cosmology and Astrophysics, College of Sciences, Northeastern University, Shenyang 110819, China}
\affiliation{MOE Key Laboratory of Data Analytics and Optimization for Smart Industry, Northeastern University, Shenyang 110819, China}
\affiliation{National Frontiers Science Center for Industrial Intelligence and Systems Optimization, Northeastern University, Shenyang 110819, China}

\begin{abstract}

Recent DESI DR2 data has shown a significant preference for dynamical dark energy, yet this has further exacerbated the $H_0$ tension. In this work, we explore the potential of interacting dark energy models ($\widetilde{\Lambda}$CDM and $e\widetilde{\Lambda}$CDM) within the asymptotic-safety framework of quantum gravitational field theory to alleviate the $H_0$ tension. We perform observational constraints using the latest baryon acoustic oscillation data from DESI DR2, cosmic microwave background (CMB) data from Planck and ACT, and type Ia supernova data from DESY5 and PantheonPlus, as well as the SH0ES data. From our analysis, we observe the dynamical scale parameter of the cosmological constant, $\delta_{\Lambda} = -0.270\pm 0.100$, in the $e\widetilde{\Lambda}$CDM model using the CMB+DESI+SH0ES data, which deviates from $\Lambda$CDM at the $2.7\sigma$ level. Simultaneously, we find $H_0 = 70.84\pm 0.74~\mathrm{km\,s^{-1}\,Mpc^{-1}}$, reducing the $H_0$ tension to $1.7\sigma$. This increase in the inferred $H_0$ is due to the anti-correlation between $\delta_{\Lambda}$ and $H_0$, whereby a negative $\delta_{\Lambda}$ leads to a higher $H_0$ value. Furthermore, for the CMB+DESI+SH0ES combination, we obtain $\Delta\chi^2_{\min}=-14.14$ and $\Delta\mathrm{DIC}=-9.18$, favoring the $e\widetilde{\Lambda}$CDM model over $\Lambda$CDM. Overall, the $e\widetilde{\Lambda}$CDM model can improve the fit and ease the $H_0$ tension, especially for the data combinations that provide the strongest statistical support.

\end{abstract}

\maketitle

\section{Introduction} 
Over the past few decades, the standard cosmological model, $\Lambda$ cold dark matter ($\Lambda$CDM), has served as the prevailing paradigm in cosmology. It provides the simplest and most direct description of our universe, successfully explaining a wide range of cosmological observations, such as cosmic microwave background (CMB)~\cite{WMAP:2003elm,Planck:2018vyg}, type Ia supernova (SN)~\cite{SupernovaSearchTeam:1998fmf,SupernovaCosmologyProject:1998vns}, and baryon acoustic oscillation (BAO)~\cite{SDSS:2005xqv,BOSS:2016wmc,eBOSS:2020yzd}. Nevertheless, despite its many achievements, the $\Lambda$CDM model still faces a number of challenges. In particular, the cosmological constant $\Lambda$, interpreted as the vacuum energy density, suffers from two fundamental theoretical issues, ``cosmic coincidence'' and ``fine-tuning'' problems~\cite{Sahni:1999gb,Bean:2005ru}. With the continuous improvement in the precision of cosmological observations, several tensions have emerged that require resolution, including the $S_8$ tension\footnote{The so-called $S_{8}$ tension refers to around $3\sigma$ discrepancy between the values of $S_{8}$ inferred from weak gravitational lensing of the Kilo-Degree Survey (KiDS)~\cite{KiDS:2020suj} and those derived from Planck data~\cite{Planck:2018vyg}. However, the updated cosmic shear analysis from KiDS-Legacy has significantly reduced this discrepancy to $0.73\sigma$~\cite{Wright:2025xka}, bringing the results into full consistency with Planck.}~\cite{DiValentino:2020vvd}, and most prominently, the $H_0$ tension~\cite{Verde:2019ivm,Riess:2019qba,Riess:2021jrx}.

Specifically, the $H_0$ tension arises from the discrepancy between the value of the Hubble constant $H_0$, inferred within the $\Lambda$CDM framework from Planck CMB data, $H_0 = 67.36 \pm 0.54 \,\mathrm{km\,s^{-1}\,Mpc^{-1}}$~\cite{Planck:2018vyg}, and the substantially higher value obtained from SH0ES, $H_0 = 73.04 \pm 1.04 \,\mathrm{km\,s^{-1}\,Mpc^{-1}}$~\cite{Riess:2021jrx}. The discrepancy between these two determinations has now reached a statistical significance exceeding $5\sigma$ level. In particular, the recent joint analysis of ground-based CMB experiments, South Pole Telescope~\cite{SPT-3G:2025bzu,SPT-3G:2025vyw} and Data Release 6 (DR6) of Atacama Cosmology Telescope (ACT)~\cite{AtacamaCosmologyTelescope:2025blo,ACT:2025llb}, has further increased the significance of the $H_0$ tension to the $6.4\sigma$ level~\cite{SPT-3G:2025bzu}. In recent years, the $H_0$ tension has sparked extensive discussion; (see Refs.~\cite{Bernal:2016gxb,Guo:2018ans,Zhao:2019gyk,Vagnozzi:2019ezj,Cai:2021wgv,Vagnozzi:2021gjh,Vagnozzi:2021tjv,Yang:2021eud,Escudero:2022rbq,James:2022dcx,Jin:2022qnj,Vagnozzi:2023nrq,Song:2022siz,Pierra:2023deu,Jin:2023sfc,Breuval:2024lsv,Huang:2024erq,Zhang:2024rra,Song:2025ddm,Jin:2025dvf,Pantos:2026cxv,Cai:2026swf}; see also Refs.~\cite{DiValentino:2021izs,Kamionkowski:2022pkx} for relevant reviews). Although considerable effort has been devoted to ruling out unknown systematic errors~\cite{Efstathiou:2020wxn,Mortsell:2021nzg,Mortsell:2021tcx,Bhardwaj:2023mau,Sharon:2023ioz}, the persistence of this tension strongly suggests the necessity of considering new physics beyond the standard cosmological model.

Numerous studies have proposed various extensions of the $\Lambda$CDM paradigm in order to address the $H_0$ tension, such as early dark energy~\cite{Poulin:2018cxd,Smith:2020rxx,Yin:2023srb,Du:2026qtq}, phenomenologically emergent dark energy~\cite{Li:2019yem,Pan:2019hac,DiValentino:2021rjj,Yao:2023ybs}, holographic dark energy~\cite{Li:2004rb,Huang:2004wt,Wang:2004nqa,Zhang:2005hs,Zhang:2005yz,Ma:2007av,Zhang:2007sh,Zhang:2009un,Landim:2015hqa,Zhang:2015rha,Feng:2016djj,Wang:2016och,Drepanou:2021jiv,Han:2024sxm,Li:2024qus}, interacting dark energy~\cite{Zhang:2004gc,Cai:2004dk,Zhang:2005rg,Zhang:2005rj,Wang:2006qw,Zhang:2006qu,Fu:2011ab,Zhang:2012uu,Zhang:2013lea,Li:2014cee,Li:2015vla,Cai:2015emx,Koyama:2015vza,Wang:2016lxa,Costa:2016tpb,DiValentino:2017iww,Yang:2018euj,Guo:2018gyo,Pan:2019gop,Feng:2017usu,Pan:2020zza,DiValentino:2019ffd,Yao:2020pji,Wang:2021kxc,Yao:2022kub,Li:2023fdk,Li:2023gtu,Giare:2024ytc,Wang:2024vmw,Li:2024qso,vanderWesthuizen:2025vcb,vanderWesthuizen:2025rip,vanderWesthuizen:2025mnw,Li:2025owk,Li:2025muv}, modified gravity~\cite{Clifton:2011jh,Joyce:2014kja,Nojiri:2017ncd}, and other possibilities~\cite{Zhang:2007bi,Cui:2009ns,Zhang:2014nta,Xue:2014kna,Guo:2015gpa,Jedamzik:2020krr,Zumalacarregui:2020cjh,Du:2025csv}. In particular, \citet{Xue:2014kna} focused on the asymptotic safety of gravitational field theory, a framework of quantum gravity, which leads to a phenomenological model of interacting dark energy. That work goes beyond the core assumption of the standard $\Lambda$CDM model that the gravitational constant $G$ and the cosmological constant $\Lambda$ are fixed. It shows that with cosmic expansion, or equivalently with redshift or energy scale evolution, $G$ and $\Lambda$ are not constant. Instead, they follow specific scaling laws and dynamically approach the present observational values $(G_0, \Lambda_0)$. The study further points out that the dynamical evolution of $G$ and $\Lambda$ originates from a nontrivial ultraviolet fixed point of quantum gravity, and that the two are intrinsically connected through critical exponents. These results provide a first-principles basis for modifying the evolution of matter, radiation, and dark energy, and they open a new theoretical direction for addressing challenges such as the $H_0$ tension.
 
Based on~\citet{Xue:2014kna} previous work, \citet{Gao:2021xnk} developed extended cosmological models motivated by the dynamical behavior of $G$ and $\Lambda$. These models introduce interactions between vacuum energy and matter or radiation, modifying the evolution of the matter or radiation density and the cosmological constant. Specifically, two parameters $\delta_{\mathrm{G}}$ and $\delta_{\Lambda}$ are introduced. The two-parameter $e\widetilde{\Lambda}$CDM model treats $\delta_{\mathrm{G}}$ and $\delta_{\Lambda}$ as independent. Another extension considers the case where $\delta_{\mathrm{G}}$ and $\delta_{\Lambda}$ are related in the low-redshift limit, which reduces to a single-parameter $\widetilde{\Lambda}$CDM model. Interestingly, \citet{Gao:2021xnk} found that the $e\widetilde{\Lambda}$CDM model performs better than the $\Lambda$CDM model in alleviating the $H_0$ tension. Therefore, it is found that the $\widetilde{\Lambda}$CDM and $e\widetilde{\Lambda}$CDM models have the potential to alleviate the $H_0$ tension, and they deserve further discussion. 

Recently, the second data release (DR2) of Dark Energy Spectroscopic Instrument (DESI)~\cite{DESI:2025zgx} has provided BAO measurements from galaxies, quasars, and the Lyman-$\alpha$ (Ly$\alpha$) forest. The combination of the BAO data from DESI DR2, SN data from DESY5, and CMB data from Planck and ACT suggests the possible dynamical evolution of dark energy, with a statistical significance of up to $4.2\sigma$. This has triggered intense discussions about possible new physics~\cite{Colgain:2024mtg,Li:2024qso,Escamilla:2024ahl,Wang:2024dka,RoyChoudhury:2024wri,Sabogal:2024yha,Giare:2024gpk,Li:2024qus,Wang:2024hwd,Jiang:2024xnu,Du:2024pai,Ye:2024ywg,Wu:2024faw,Dinda:2024ktd,Li:2024bwr,Huang:2025som,Barua:2025ypw,Yashiki:2025loj,Ling:2025lmw,Goswami:2025uih,Yang:2025boq,Ozulker:2025ehg,Cheng:2025lod,Liu:2025myr,Gialamas:2025pwv,Yang:2025ume,Chen:2025wwn,Kumar:2025etf,Abedin:2025dis,Araya:2025rqz,Li:2025dwz,Paliathanasis:2025xxm,Du:2025iow,RoyChoudhury:2025dhe,Cai:2025mas,Li:2025cxn,Avila:2025sjz,FrancoAbellan:2025fkb,Paliathanasis:2025kmg,Li:2025htp,Hussain:2025uye,Samanta:2025oqz,Yang:2025oax,Paul:2025wix,GarciaEscudero:2025lef,Zhou:2025nkb,Feng:2025mlo,Petri:2025swg,vanderWesthuizen:2025iam,Li:2025eqh,Du:2025xes,Wu:2025vfs,Li:2025owk,Pedrotti:2025ccw,Yao:2025kuz,Yadav:2025vgo,Alam:2025epg,Jia:2025poj,Wang:2025vtw,Chaudhary:2025pcc,Hogas:2025ahb,Colgain:2024xqj,Yadav:2025vpx,Li:2025vuh,Li:2026xaz,Li:2026ldf,Hou:2026phk,Feng:2026pzs}. Furthermore, several recent studies further intensify the $H_0$ tension in light of DESI DR2~\cite{Pang:2025lvh,Ye:2025ark}. At the same time, it remains premature to conclude whether the DESI results truly support dynamical dark energy or instead point toward a new physical paradigm. The reported preference of DESI for dynamical dark energy can also be interpreted within alternative frameworks, such as interacting dark energy models~\cite{Li:2024qso,Giare:2024smz,Li:2025owk,Li:2025ula,Yang:2025uyv,Shah:2025ayl,Silva:2025hxw,Pan:2025qwy,Wang:2025znm,Li:2025muv}. In this context, it is essential to revisit the $\widetilde{\Lambda}$CDM and $e\widetilde{\Lambda}$CDM models, as well as their potential to alleviate the $H_0$ tension. 

In this work, we utilize the latest DESI DR2 BAO measurements, CMB data from Planck and ACT DR6, and the SN compilations from DESY5 and PantheonPlus to constrain the $\widetilde{\Lambda}$CDM and $e\widetilde{\Lambda}$CDM models. Our motivation is to explore the ability of these interacting models to alleviate the $H_0$ tension in light of DESI, and to provide updated constraints on them with the new data. 


\section{Methodology and data}\label{sec2}
We consider a spatially flat, homogeneous, and isotropic universe described by the Friedmann-Robertson-Walker (FRW) metric within the framework of general relativity. The Hubble expansion rate satisfies
\begin{equation}
H^2(z) = \frac{8\pi G(z)}{3}\left[\rho_{\mathrm{m}}(z)+\rho_{\mathrm{r}}(z)+\rho_{\Lambda}(z)\right],
\end{equation}
where $\rho_{\mathrm{m}}$, $\rho_{\mathrm{r}}$, and $\rho_{\Lambda}$ denote the matter, radiation, and dark energy densities, respectively. 
To account for quantum gravity effects, we introduce the scale-dependent couplings $G(z)$ and $\Lambda(z)$ following the asymptotic safety scenario~\cite{Weinberg:2009wa,Xue:2014kna,Xue:2020nzb,Gao:2021xnk},
\begin{align}
\frac{G(z)}{G_0}=(1+z)^{-\delta_G}, \quad
\frac{\Lambda(z)}{\Lambda_0}=(1+z)^{\delta_\Lambda},
\end{align}
where $\delta_G, \delta_\Lambda \ll 1$ are critical indices characterizing their mild redshift dependence. The detailed derivation of the modified background evolution and the relation between the critical indices are given in Appendix~\ref{appendixA}. The modified Friedmann equation becomes
\begin{align}
\label{eq:2.10}
E^2(z)&=\Omega_{\mathrm{m0}}(1+z)^{3-\delta_G}
+\Omega_{\mathrm{r0}}(1+z)^{4-\delta_G}
+\Omega_{\Lambda0}(1+z)^{\delta_\Lambda},
\end{align}
with $E(z)\equiv H(z)/H_0$ and $\Omega_{\mathrm{m0}}+\Omega_{\mathrm{r0}}+\Omega_{\Lambda0}=1$.  
Consistency with the generalized conservation law yields a low-redshift relation between the two critical indices,
\begin{equation}
\delta_\Lambda \simeq 0.47\,\delta_G,
\end{equation}
leading to a one-parameter extension of the $\Lambda$CDM model (denoted $\widetilde{\Lambda}$CDM), while treating $\delta_G$ and $\delta_\Lambda$ as independent defines a two-parameter $e\widetilde{\Lambda}$CDM extension.  
For comparison, the dynamical dark-energy $w_0w_a$CDM model adopts
\begin{equation}
w(z)=w_0+w_a\frac{z}{1+z},
\end{equation}
with
\begin{equation}
\begin{aligned}
H(z) = H_0 \biggl[ &\Omega_{\mathrm{m0}}(1+z)^3 + \\
                   &\Omega_{\mathrm{de0}}(1+z)^{3(1+w_0+w_a)}e^{-3w_a z/(1+z)} \biggr]^{1/2}.
\end{aligned}
\end{equation}

\begin{table}[t]
\centering
\renewcommand{\arraystretch}{1.2}
\caption{Flat priors on the main cosmological parameters for different models.}
\label{tab:priors_models}

\begingroup
\small
\setlength{\extrarowheight}{0.5ex} 
\renewcommand{\arraystretch}{1.25} 

\begin{tabular*}{\linewidth}{@{\hspace{0.4cm}} @{\extracolsep{\fill}} c c c @{\hspace{0.4cm}}}
\hline\hline
\textbf{Model} & \textbf{Parameter} & \textbf{Prior} \\
\hline
\multirow{6}{*}{\textbf{$\Lambda$CDM}} 
  & $\Omega_{\mathrm{b}} h^2$ & $\mathcal{U}[0.05, 0.1]$ \\
  & $\Omega_{\mathrm{c}} h^2$ & $\mathcal{U}[0.01, 0.99]$ \\
  & $100\theta_{\mathrm{MC}}$ & $\mathcal{U}[0.5, 10]$ \\
  & $\tau_{\mathrm{reio}}$ & $\mathcal{U}[0.01, 0.8]$ \\
  & $\ln 10^{10}A_{\mathrm{s}}$ & $\mathcal{U}[1.61, 3.91]$ \\
  & $n_{\mathrm{s}}$ & $\mathcal{U}[0.8, 1.2]$ \\
\hline
\multirow{2}{*}{\textbf{$w_{0}w_{a}$CDM}} 
  & $w_0$ & $\mathcal{U}[-3, 1]$ \\
  & $w_a$ & $\mathcal{U}[-3, 1]$ \\
\hline
\raisebox{-0.4ex}{\textbf{$\widetilde{\Lambda}$CDM}} & $\delta_{\mathrm{G}}$ & $\mathcal{U}[-1, 1]$ \\
\hline
\multirow{2}{*}{\raisebox{-0.4ex}{\textbf{$e\widetilde{\Lambda}$CDM}}} 
  & $\delta_{\mathrm{G}}$ & $\mathcal{U}[-1, 1]$ \\
  & $\delta_{\Lambda}$ & $\mathcal{U}[-1, 1]$ \\
\hline\hline
\end{tabular*}
\endgroup
\end{table}

\begin{table*}[!htbp]
\centering
\caption{Cosmological parameter constraints ($1\sigma$ confidence level) for $\Lambda$CDM, ${w_0w_a}$CDM, $\widetilde{\Lambda}$CDM, and $e\widetilde{\Lambda}$CDM models from the CMB+DESI, CMB+DESI+DESY5, and CMB+DESI+PantheonPlus datasets. Here, $H_{0}$ is in units of ${\rm km}~{\rm s}^{-1}~{\rm Mpc}^{-1}$.}
\label{tab:combined_params}
\renewcommand{\arraystretch}{1.6}
\scriptsize 

\resizebox{\textwidth}{!}{
\begin{tabular}{lccccccc}
\hline\hline
\textbf{Model / Dataset} & $H_0$ & $\Omega_{\mathrm{m}}$ &  $\delta_{\mathrm{G}}$ & $\delta_{\Lambda}$ & $w_0$ & $w_a$ \\
\hline

\multicolumn{7}{l}{\textbf{$\Lambda$CDM}} \\
CMB+DESI              & $68.15\pm 0.28$ & $0.3031\pm 0.0036$  & $-$ & $-$ & $-$ & $-$ \\
CMB+DESI+DESY5        & $67.98\pm 0.27$         & $0.3053\pm 0.0036$  & $-$ & $-$ & $-$ & $-$ \\
CMB+DESI+PantheonPlus & $68.08\pm 0.27$         & $0.3039\pm 0.0035$  & $-$ & $-$ & $-$ & $-$ \\
\hline

\multicolumn{7}{l}{\textbf{${w_0w_a}\text{CDM}$}} \\
CMB+DESI              & $63.90^{+1.70}_{-2.10}$ & $0.3500\pm 0.0210$ & $-$ & $-$ & $-0.450\pm 0.210$ & $-1.67\pm 0.59$ \\
CMB+DESI+DESY5        & $66.78\pm 0.57$         & $0.3187\pm 0.0056$ & $-$ & $-$ & $-0.753\pm 0.055$ & $-0.86\pm 0.21$ \\
CMB+DESI+PantheonPlus & $67.55\pm 0.60$         & $0.3111\pm 0.0057$ & $-$ & $-$ & $-0.838\pm 0.053$ & $-0.62\pm 0.20$ \\
\hline

\multicolumn{7}{l}{\textbf{$\widetilde{\Lambda}$CDM}} \\
CMB+DESI              & $68.59\pm 0.38$         & $0.2998\pm 0.0040$  & $-0.00087\pm 0.00051$ & $-$ & $-$ & $-$ \\
CMB+DESI+DESY5        & $68.30\pm 0.35$         & $0.3030\pm 0.0038$  & $-0.00066^{+0.00053}_{-0.00046}$ & $-$ & $-$ & $-$ \\
CMB+DESI+PantheonPlus & $68.47\pm 0.36$         & $0.3011\pm 0.0038$  & $ -0.00075\pm 0.00049$ & $-$ & $-$ & $-$ \\
\hline

\multicolumn{7}{l}{\textbf{$e\widetilde{\Lambda}$CDM}} \\
CMB+DESI              & $69.01\pm 0.91$         & $0.2967\pm 0.0073$  & $-0.00074\pm 0.00056$ & $-0.060\pm 0.1200$ & $-$ & $-$ \\
CMB+DESI+DESY5        & $67.41\pm 0.56$ & $0.3093\pm 0.0051$  & $-0.00107\pm 0.00055$ & $0.149\pm 0.0694$ & $-$ & $-$ \\
CMB+DESI+PantheonPlus & $67.98\pm 0.60$         & $0.3048\pm 0.0053$  & $-0.00096\pm 0.00053$ & $0.075\pm 0.0750$ & $-$ & $-$ \\
\hline\hline
\end{tabular}
}
\end{table*}

We employ a modified version of the Boltzmann solver \texttt{CAMB}\footnote{\url{https://github.com/liaocrane/IDECAMB}.}~\cite{Lewis:1999bs,Li:2023fdk} to implement extensions of the standard cosmological model. The parameter estimation is performed via Markov Chain Monte Carlo (MCMC) analysis using the Bayesian inference framework \texttt{Cobaya}\footnote{\url{https://github.com/CobayaSampler/cobaya}.}~\cite{Torrado:2020dgo}, which interfaces efficiently with Boltzmann solvers and samplers. Convergence of the MCMC chains is ensured by the Gelman-Rubin criterion $R-1 < 0.02$~\cite{Gelman:1992zz}, and the resulting chains are analyzed using \texttt{GetDist}\footnote{\url{https://github.com/cmbant/getdist}.}~\cite{Lewis:2019xzd}. The free parameters and uniform priors adopted for each model are summarized in Table~\ref{tab:priors_models}. Our baseline datasets include the CMB data from Planck~\cite{Planck:2018vyg,Efstathiou:2019mdh,Planck:2019nip,Rosenberg:2022sdy} and ACT DR6~\cite{ACT:2023dou}, BAO data from DESI DR2~\cite{DESI:2025zgx}, and SN data from DESY5~\cite{DES:2024jxu} and PantheonPlus~\cite{Brout:2022vxf}. Moreover, we also used a Gaussian prior, $H_0 = 73.04 \pm 1.04 \,\mathrm{km\,s^{-1}\,Mpc^{-1}}$ from SH0ES~\cite{Riess:2021jrx}.

\begin{figure*}[htbp]
\centering
\includegraphics[width=0.46\textwidth, height=0.305\textheight]{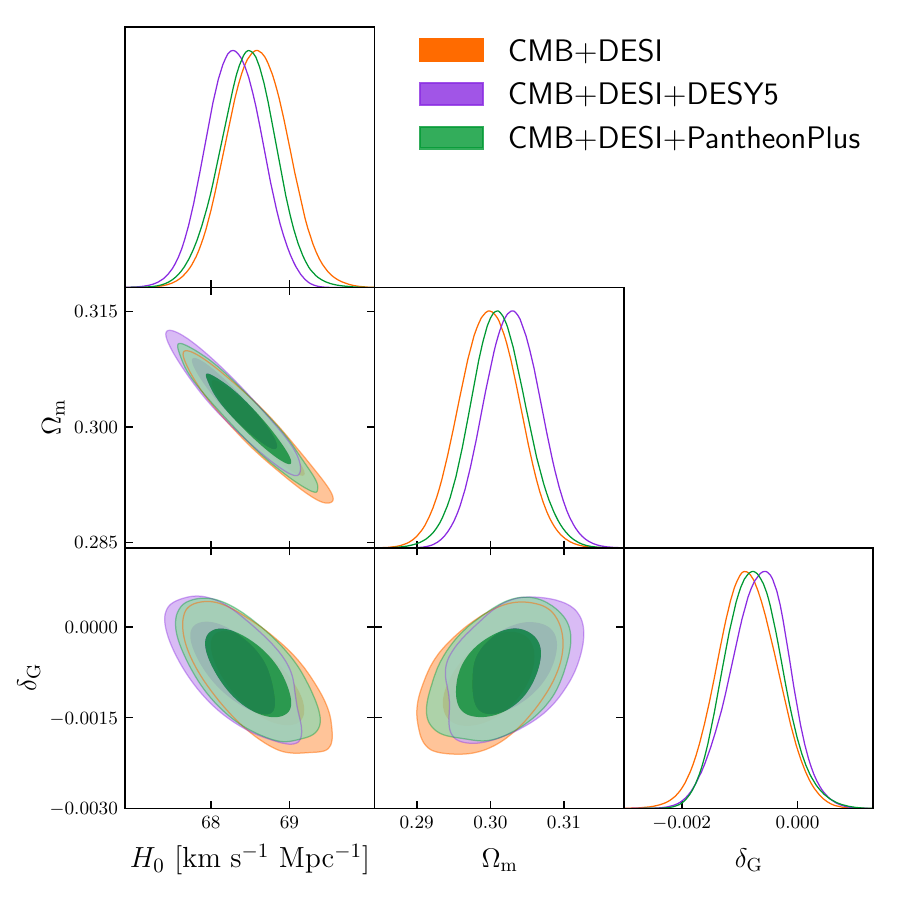} \hspace{1cm}
\includegraphics[width=0.44\textwidth, height=0.3\textheight]{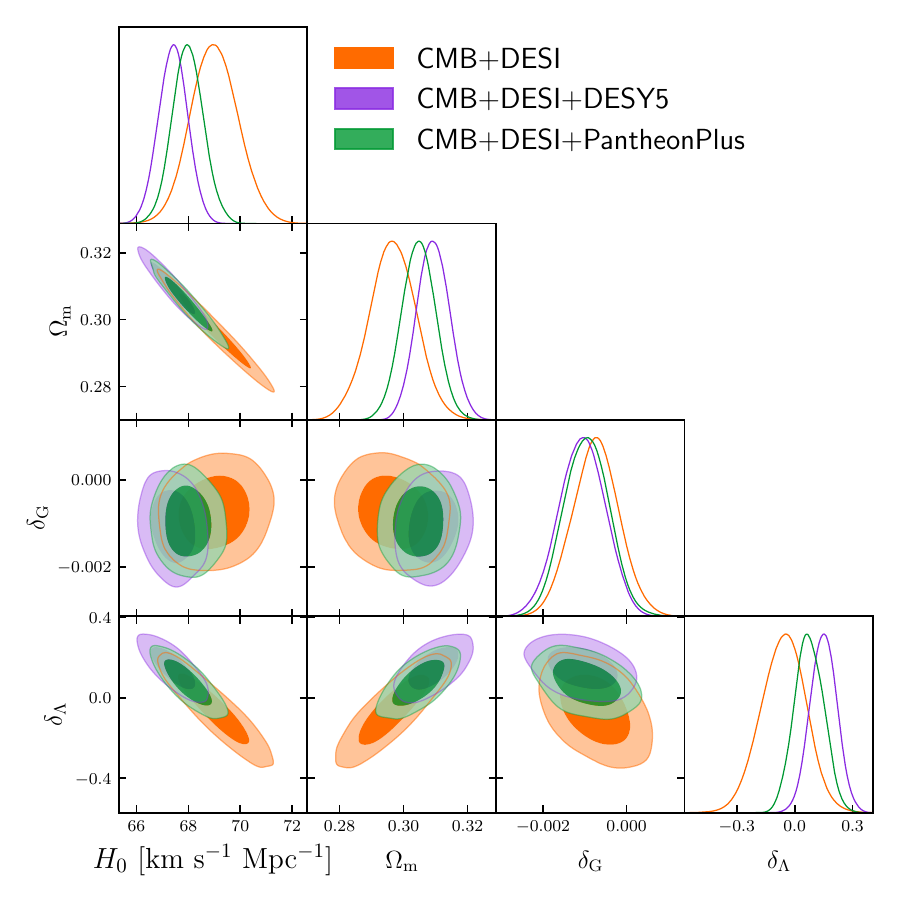}
\caption{\label{fig1} Constraints on the cosmological parameters from the combination of CMB, BAO, and SN data. \textit{Left panel:} Constraints on the cosmological parameters obtained using the CMB+DESI, CMB+DESI+DESY5, and CMB+DESI+PantheonPlus data in the $\widetilde{\Lambda}$CDM model. \textit{Right panel:} Constraints on the cosmological parameters obtained using the CMB+DESI, CMB+DESI+DESY5, and CMB+DESI+PantheonPlus data in the $e\widetilde{\Lambda}$CDM model.
 }
\end{figure*}

\begin{table*}[!htbp]
\centering
\caption{Fitting results ($1\sigma$ confidence level) in the $\widetilde{\Lambda}$CDM and $e\widetilde{\Lambda}$CDM models from the CMB+DESI+SH0ES, CMB+DESI+DESY5+SH0ES, and CMB+DESI+PantheonPlus+SH0ES data. Here, $H_{0}$ is in units of ${\rm km}~{\rm s}^{-1}~{\rm Mpc}^{-1}$.}
\label{tab:with_shoes} 
\setlength{\tabcolsep}{1.3mm}  
\renewcommand{\arraystretch}{1.9}  
\footnotesize 

\begin{tabular}{lcccc}
\hline\hline  
\textbf{Model / Dataset} & $H_0$ & $\Omega_{\mathrm{m}}$ & $\delta_{\mathrm{G}}$ & $\delta_{\Lambda}$ \\
\hline 

\multicolumn{5}{l}{\textbf{$\widetilde{\Lambda}$CDM }} \\ 
CMB+DESI+SH0ES               & $69.11\pm 0.34$         & $0.2947\pm 0.0036$         & $-0.00130\pm 0.00048$         & $-$  \\
CMB+DESI+DESY5+SH0ES         & $68.85\pm 0.34$         & $0.2976\pm 0.0036$         & $-0.00115\pm 0.00050$         & $-$  \\
CMB+DESI+PantheonPlus+SH0ES  & $68.98\pm 0.37$         & $0.2962\pm 0.0038$         & $-0.00123\pm 0.00052$         & $-$  \\
\hline  

\multicolumn{5}{l}{\textbf{$e\widetilde{\Lambda}$CDM }} \\  
CMB +DESI+SH0ES               & $70.84\pm 0.74$         & $0.2827\pm 0.0056$         & $-0.00061\pm 0.00054$         & $-0.270\pm 0.100$  \\
CMB+DESI +DESY5 +SH0ES         & $68.68\pm 0.50$         & $0.2985\pm 0.0044$         & $-0.00129\pm 0.00053$         & $0.036\pm 0.064$  \\
CMB+DESI+PantheonPlus +SH0ES  & $69.20\pm 0.53$         & $0.2946\pm 0.0044$         & $-0.00111\pm 0.00053$         & $-0.038\pm 0.069$  \\
\hline\hline 
\end{tabular}
\end{table*}

\begin{figure*}[htbp]
\centering
\includegraphics[width=0.47\textwidth, height=0.305\textheight]{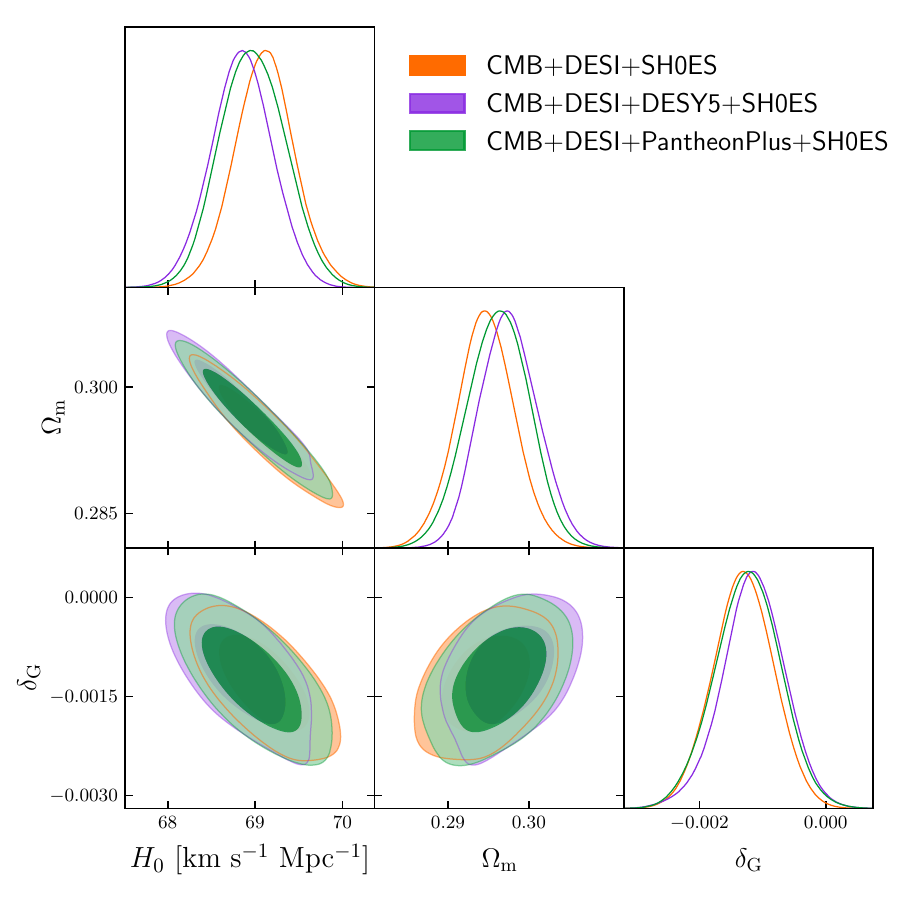} \hspace{1cm}
\includegraphics[width=0.45\textwidth, height=0.3\textheight]{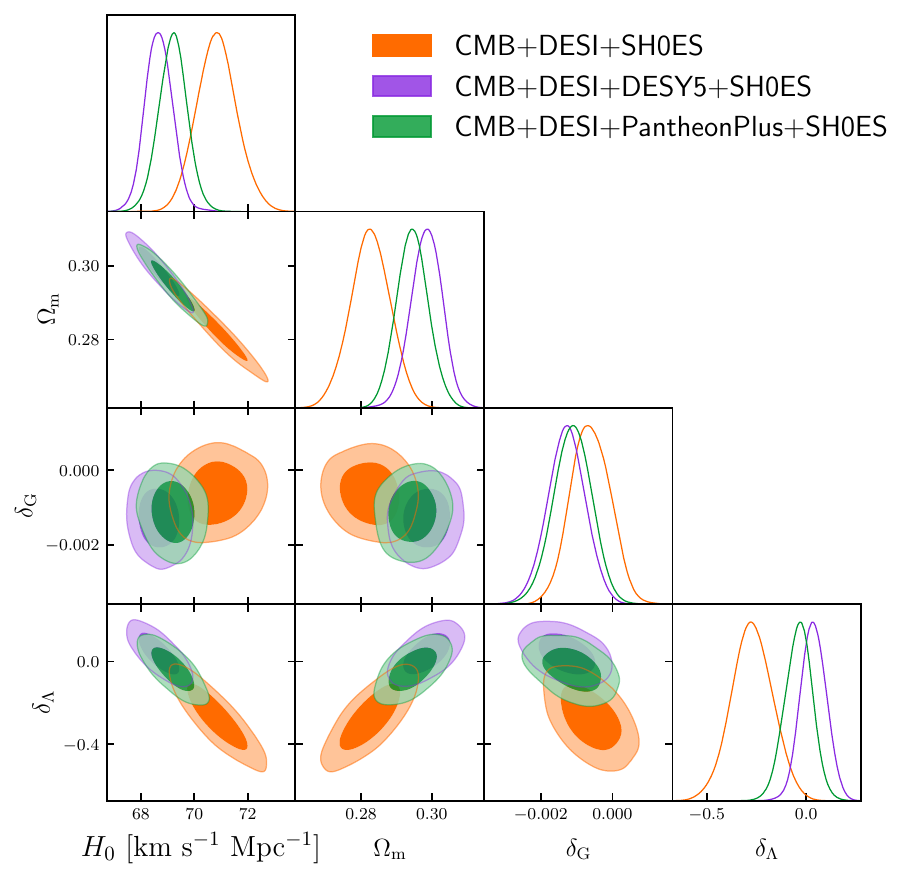}
\caption{\label{fig2} Constraints on the cosmological parameters from the combination of CMB, BAO, SN, and SH0ES data. \textit{Left panel:} Constraints on the cosmological parameters obtained using the CMB+DESI+SH0ES, CMB+DESI+DESY5+SH0ES, and CMB+DESI+PantheonPlus+SH0ES data in the $\widetilde{\Lambda}$CDM model. \textit{Right panel:} Constraints on the cosmological parameters obtained using the CMB+DESI+SH0ES, CMB+DESI+DESY5+SH0ES, and CMB+DESI+PantheonPlus+SH0ES data in the $e\widetilde{\Lambda}$CDM model.
}
\end{figure*}

\begin{table}
\centering
\caption{Comparison of $\Delta\chi^2_{\min}$ and $\Delta\mathrm{DIC}$ for $\widetilde{\Lambda}$CDM and $e\widetilde{\Lambda}$CDM relative to $\Lambda$CDM. Negative values indicate that the corresponding extended model is favored over $\Lambda$CDM according to that statistic.}
\label{tab:ic_comparison}
\setlength{\tabcolsep}{1.3mm}
\renewcommand{\arraystretch}{1.6}
\footnotesize
\begin{tabular}{lcc}
\hline\hline
\textbf{Model / Dataset} & $\Delta \chi^2_{\min}$ & $\Delta\mathrm{DIC}$ \\
\hline
\multicolumn{3}{l}{\textbf{$\widetilde{\Lambda}$CDM}} \\
CMB+DESI               & $-1.33$  & $-1.01$ \\
CMB+DESI+DESY5         & $-1.44$  & $0.18$  \\
CMB+DESI+PantheonPlus  & $-0.60$  & $-0.02$ \\
CMB+DESI+SH0ES                & $-3.18$  & $-3.77$ \\
CMB+DESI+DESY5+SH0ES          & $-2.38$  & $-1.85$ \\
CMB+DESI+PantheonPlus+SH0ES   & $-3.20$  & $-0.58$ \\
\hline
\multicolumn{3}{l}{\textbf{$e\widetilde{\Lambda}$CDM}} \\
CMB+DESI               & $-3.45$  & $0.33$  \\
CMB+DESI+DESY5         & $-7.76$  & $-3.66$ \\
CMB+DESI+PantheonPlus  & $-4.64$  & $0.37$  \\
CMB+DESI+SH0ES                & $-14.14$ & $-9.18$ \\
CMB+DESI+DESY5+SH0ES          & $-6.76$  & $-3.73$ \\
CMB+DESI+PantheonPlus+SH0ES   & $-6.52$  & $-1.27$ \\
\hline\hline
\end{tabular}
\end{table}

\section{Results and discussions}
\begin{figure*}[htbp]
\includegraphics[scale=0.35]{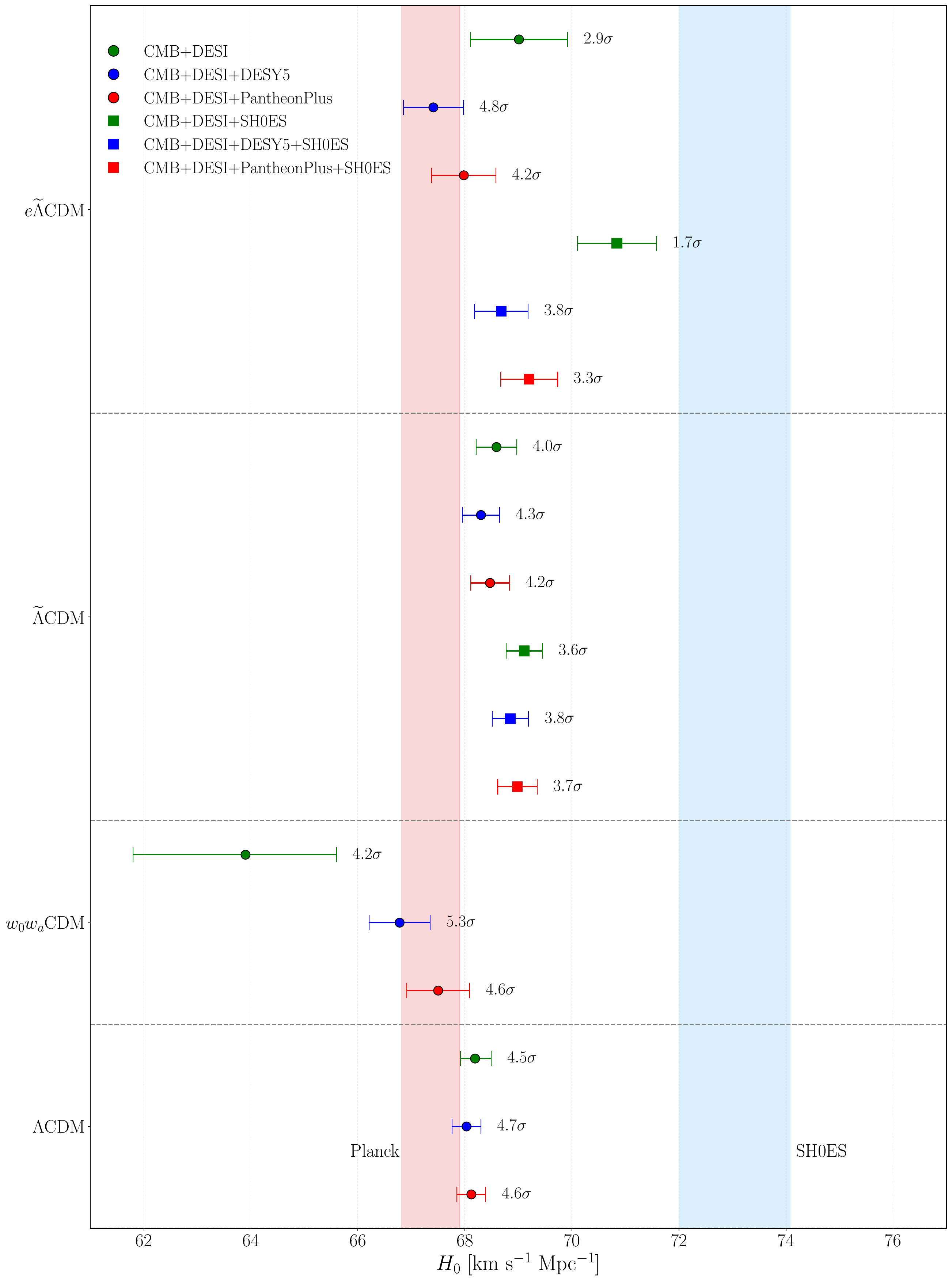}
\centering
\caption{\label{fig3} The $1\sigma$ error bar plots of $H_{0}$ in various cosmological models from the CMB+DESI, CMB+DESI+DESY5, CMB+DESI+PantheonPlus, CMB+DESI+SH0ES, CMB+DESI+DESY5+SH0ES, and CMB+DESI+PantheonPlus+SH0ES data. The circular markers represent results without including the SH0ES prior, while the square markers denote the constraints obtained when the SH0ES data are included.}
\end{figure*}

In this section, we present and analyze the constraints on cosmological parameters. We consider the $\Lambda$CDM, $w_{0}w_{a}$CDM, $\widetilde{\Lambda}$CDM, and $e\widetilde{\Lambda}$CDM models using the current observational data, including CMB, DESI, DESY5, PantheonPlus, and SH0ES, for cosmological analysis. The $1\sigma$ errors for the marginalized parameter constraints are summarized in Tables~\ref{tab:combined_params} and~\ref{tab:with_shoes}. We show the $1\sigma$ and $2\sigma$ posterior distribution contours for the cosmological parameters in the $\widetilde{\Lambda}$CDM and $e\widetilde{\Lambda}$CDM models in Figs.~\ref{fig1} and~\ref{fig2}. We present the error bar plots of $H_{0}$ at the $1\sigma$ confidence level in Fig.~\ref{fig3}. In addition, Table~\ref{tab:ic_comparison} summarizes the values of $\Delta\chi^2_{\min}$ and $\Delta\mathrm{DIC}$ for the $\widetilde{\Lambda}$CDM and $e\widetilde{\Lambda}$CDM models relative to $\Lambda$CDM.

In the left panel of Fig.~\ref{fig1}, we display the constraints on the $\widetilde{\Lambda}$CDM model obtained from CMB and DESI BAO combined with SN data. The constraints on the parameter $\delta_{\mathrm{G}}$ are $-0.00087\pm 0.00051$, $-0.00066^{+0.00053}_{-0.00046}$, and $-0.00075\pm 0.00049$, corresponding to $1.7\sigma$, $1.3\sigma$, and $1.1\sigma$ deviations from zero, using CMB+DESI, CMB+DESI+DESY5, CMB+DESI+PantheonPlus, respectively. A negative $\delta_{\mathrm{G}}$ implies that in the $\widetilde{\Lambda}$CDM framework, the effective cosmological constant grows with time, producing a phantom-like behavior that enhances cosmic acceleration. Consequently, an anti-correlation between $H_{0}$ and $\delta_{\mathrm{G}}$ is evident. With CMB+DESI alone, we obtain a more negative $\delta_{\mathrm{G}}$, resulting in a slightly higher value of $H_{0}=68.59 \pm 0.38 \,\mathrm{km\,s^{-1}\,Mpc^{-1}}$ than in the $\Lambda$CDM model. 


In the right panel of Fig.~\ref{fig1}, we display the constraints on the $e\widetilde{\Lambda}$CDM model obtained from CMB and DESI BAO combined with SN data. For CMB+DESI alone, the constraint on $\delta_{\mathrm{G}}$ is $\delta_{\mathrm{G}} = -0.00074 \pm 0.00056$, consistent with that in $\widetilde{\Lambda}$CDM. The result for $\delta_{\Lambda}$ is $-0.06\pm 0.1200$, which corresponds to $0.5\sigma$. From the right panel of Fig.~\ref{fig1}, it is clear that, since we treat $\delta_{\Lambda}$ and $\delta_{\mathrm{G}}$ as free parameters rather than fixing them to a linear relationship, $\delta_{\Lambda}$ shows a significant anti-correlation with $H_0$. Therefore, a more negative $\delta_{\Lambda}$ leads to a higher $H_0 = 69.01\pm 0.91\,\mathrm{km\,s^{-1}\,Mpc^{-1}}$, alleviating the $H_0$ tension to $2.9\sigma$. However, once SN data are included, we find that $\delta_{\Lambda}$ is driven toward positive values, which in turn lowers the inferred value of $H_0$. For instance, for CMB+DESI+DESY5 we obtain $\delta_{\Lambda} = 0.149 \pm 0.0694$, corresponding to $H_0 = 67.41 \pm 0.56\,\mathrm{km\,s^{-1}\,Mpc^{-1}}$, which is slightly lower than the $\Lambda$CDM result. A similar trend is observed for CMB+DESI+PantheonPlus, where $\delta_{\Lambda} = 0.075 \pm 0.0750$ and $H_0 = 67.98 \pm 0.60\,\mathrm{km\,s^{-1}\,Mpc^{-1}}$, which is essentially consistent with the $\Lambda$CDM model. We also note that the uncertainty in $H_0$ is slightly enlarged in the $e\widetilde{\Lambda}$CDM model, as expected for an extension with additional free parameters.

In Fig.~\ref{fig2}, we present the results for the $\widetilde{\Lambda}$CDM and $e\widetilde{\Lambda}$CDM models using the combinations CMB+DESI+SH0ES, CMB+DESI+DESY5+SH0ES, and CMB+DESI+PantheonPlus+SH0ES. For the $\widetilde{\Lambda}$CDM model, we observe that the inclusion of SH0ES leads to a slight increase in $H_0$. However, in the $e\widetilde{\Lambda}$CDM model, the inclusion of SH0ES significantly raises $H_0$. Specifically, CMB+DESI+SH0ES gives $H_0 = 70.84\pm 0.74 \,\mathrm{km\,s^{-1}\,Mpc^{-1}}$, alleviating the $H_0$ tension to $1.7\sigma$. Due to the significant anti-correlation between $H_0$ and $\delta_{\Lambda}$, this leads to a more negative $\delta_{\Lambda} = -0.270\pm 0.100$, which deviates from zero at the $2.7\sigma$ level. 

To clearly illustrate how different models mitigate the $H_{0}$ tension across various data combinations, we show the error-bar plot for $H_{0}$ in Fig.~\ref{fig3}. Notably, although the DESI data report a significant preference for the $w_{0}w_{a}$CDM model, it further exacerbates the $H_{0}$ tension relative to $\Lambda$CDM; even with CMB+DESI, the tension reaches $4.2\sigma$. In contrast, the $\widetilde{\Lambda}$CDM model provides a mild alleviation of the $H_{0}$ tension. For example, CMB+DESI and CMB+DESI+SH0ES reduce $H_{0}$ tension to $4.0\sigma$ and $3.6\sigma$, respectively. The $e\widetilde{\Lambda}$CDM model can, in some cases, substantially ease the tension; in particular, using CMB+DESI+SH0ES, the tension is reduced to $1.7\sigma$ \big($H_{0}=70.84\pm 0.74 \,\mathrm{km\,s^{-1}\,Mpc^{-1}}$\big). Although the absence of SH0ES weakens this mitigation, CMB+DESI alone still lowers the tension to $2.9\sigma$ ($H_0 = 69.01 \pm 0.91\,\mathrm{km\,s^{-1}\,Mpc^{-1}}$). A direct comparison of the best-fit DESI distance predictions in the $\Lambda$CDM, $\widetilde{\Lambda}$CDM, and $e\widetilde{\Lambda}$CDM models is presented in Appendix~\ref{appendixB}.

To quantify more rigorously the overall statistical performance of the extended models, we compare both the improvement in the best-fit likelihood and the corresponding penalty associated with the additional model complexity, always relative to $\Lambda$CDM. Specifically, for each model we compute $\Delta\chi^2_{\min}$, defined as the difference between the minimum $\chi^2$ of the model and that of $\Lambda$CDM. We also compute the DIC, which provides a complementary model-selection diagnostic beyond the best-fit likelihood alone and is widely used in posterior-based cosmological model comparison. Following the standard definition, the DIC is given by $\mathrm{DIC}=D(\bar{\theta})+2p_D$, where $D(\theta)=-2\log\mathcal{L}(\theta)$ and $p_D=\overline{D(\theta)}-D(\bar{\theta})$ measures the effective complexity of the model. Here $\overline{D(\theta)}$ denotes the posterior-averaged deviance, while $D(\bar{\theta})$ is the deviance evaluated at the posterior mean. The corresponding values of $\Delta\chi^2_{\min}$ and $\Delta\mathrm{DIC}$ relative to $\Lambda$CDM are listed in Table~\ref{tab:ic_comparison}, with negative values indicating a preference for the extended model. We find that the $\widetilde{\Lambda}$CDM model leads only to modest improvements relative to $\Lambda$CDM. By contrast, the $e\widetilde{\Lambda}$CDM model shows a more noticeable improvement in the fit for several dataset combinations. In particular, for CMB+DESI+SH0ES, we obtain $\Delta\chi^2_{\min}=-14.14$ and $\Delta\mathrm{DIC}=-9.18$, which corresponds to the strongest statistical preference among all the cases considered. This indicates that, for this dataset combination, the improved fit of the $e\widetilde{\Lambda}$CDM model remains statistically competitive even after the complexity penalty is taken into account.

\section{Conclusion}
In this work, we utilize the DESI DR2 BAO, CMB, SN, and SH0ES data to constrain the $\widetilde{\Lambda}$CDM and $e\widetilde{\Lambda}$CDM models. Our aim is to further investigate the ability of these models to ease the $H_{0}$ tension in light of the recent DESI DR2 data.

We find that when only CMB+DESI data are employed, the $e\widetilde{\Lambda}$CDM model yields $\delta_{\Lambda} = -0.06\pm 0.1200$, corresponding to a higher Hubble constant of $H_{0} = 69.01\pm 0.91~\mathrm{km\,s^{-1}\,Mpc^{-1}}$, thereby reducing the $H_{0}$ tension to $2.9\sigma$. This is because $\delta_{\Lambda}$ shows a strong anti-correlation with $H_{0}$, so that a more negative $\delta_{\Lambda}$ leads to a higher inferred $H_{0}$ value. Conversely, the $\widetilde{\Lambda}$CDM model yields a small deviation of $\delta_{\Lambda}$ from zero, indicating a limited ability to alleviate the $H_0$ tension. When using the CMB+DESI+SH0ES data, the capability of the $e\widetilde{\Lambda}$CDM model to mitigate the $H_{0}$ tension improves significantly, yielding $H_{0} = 70.84\pm 0.74~\mathrm{km\,s^{-1}\,Mpc^{-1}}$, which corresponds to a reduction of the tension to $1.7\sigma$. This result is accompanied by a more negative value of $\delta_{\Lambda} = -0.270\pm 0.100$, deviating from zero at the $2.7\sigma$ level. Moreover, for this same dataset combination, the model-comparison statistics favor the $e\widetilde{\Lambda}$CDM model, with $\Delta\chi^2_{\min}=-14.14$ and $\Delta\mathrm{DIC}=-9.18$, and the model also fits the DESI data better than $\Lambda$CDM. In fact, although DESI data have reported a clear preference for the $w_{0}w_{a}$CDM model, this preference actually exacerbates the $H_{0}$ tension relative to the $\Lambda$CDM model; even when using the combined CMB+DESI data, the tension increases to $4.2\sigma$. In contrast, the $e\widetilde{\Lambda}$CDM model exhibits a remarkable ability to reduce the $H_{0}$ tension with different datasets, with the optimal case achieving a reduction to $1.7\sigma$, highlighting its potential as a compelling alternative cosmological scenario beyond the standard $\Lambda$CDM framework. 

Overall, for some dataset combinations, the $e\widetilde{\Lambda}$CDM model, as an interacting dark energy scenario, not only exhibits a nearly $3\sigma$ deviation from $\Lambda$CDM but also demonstrates a strong capability to alleviate the $H_{0}$ tension. In the coming years, the combination of DESI BAO data with CMB and SN datasets will enhance our understanding of the nature of dark energy, further alleviate the $H_{0}$ tension, and support the existence of interactions.

\section*{Acknowledgments}
We thank Yun-He Li and Jia-Le Ling for their helpful discussions. This work was supported by the National Natural Science Foundation of China (Grants Nos. 12533001, 12575049, and 12473001), the National SKA Program of China (Grants Nos. 2022SKA0110200 and 2022SKA0110203), the China Manned Space Program (Grant No. CMS-CSST-2025-A02), and the National 111 Project (Grant No. B16009).

\textbf{Data Availability Statement} This manuscript has no associated data. [Author’s comment: All data analysed in this work is publicly available and references have been provided in Sect.~\ref{sec2}.]

\textbf{Code Availability Statement} This manuscript has no associated code/software. [Author’s comment: In this work, we utilized the publicly available software packages Cobaya and GetDist, with relevant references provided in Sect.~\ref{sec2}]

\appendix

\section{Theoretical Method}\label{appendixA}

In a spatially flat FRW universe, the redshift evolution of matter and radiation components is given by
\begin{align}
\rho_{\mathrm{m}}(z) &= \rho_{\mathrm{m0}}(1+z)^3, \\
\rho_{\mathrm{r}}(z) &= \rho_{\mathrm{r0}}(1+z)^4,
\end{align}
where $\rho_{\mathrm{m0}}$ and $\rho_{\mathrm{r0}}$ are the present-day energy densities of matter and radiation, respectively. The vacuum energy density is expressed as
\begin{equation}
\rho_{\Lambda} = \rho_{\Lambda0} = \frac{\Lambda_0}{8\pi G_0},
\end{equation}
with $\Lambda_0$ and $G_0$ denoting the current values of the cosmological constant and Newton’s gravitational constant.

\begin{figure*}[htbp]
\centering
\includegraphics[width=0.95\linewidth]{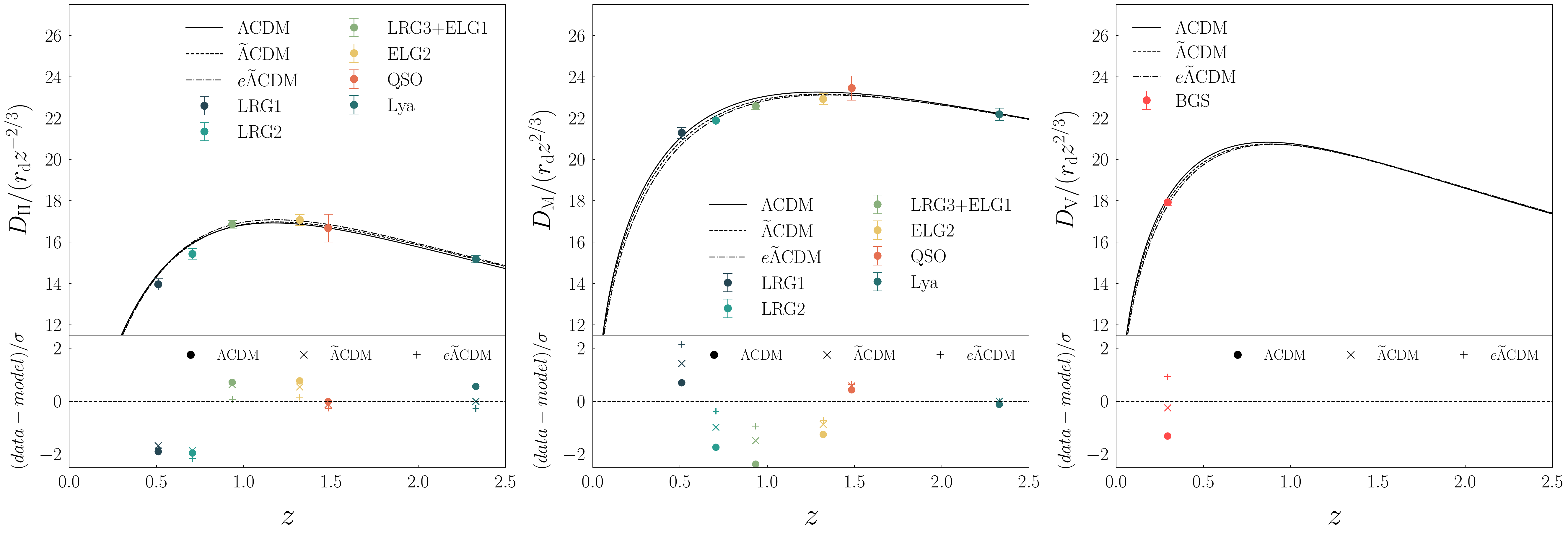}
\caption{\label{fig4} Best-fit predictions for distance-redshift relations for the $\Lambda$CDM, $\widetilde{\Lambda}$CDM, and $e\widetilde{\Lambda}$CDM models using CMB and DESI data.
\textit{Upper panel:} Best-fit predictions for distance-redshift relations for $\Lambda$CDM (solid line), $\widetilde{\Lambda}$CDM (dotted-dashed line), and $e\widetilde{\Lambda}$CDM (dashed line) obtained from the analysis of CMB+DESI data. The predictions encompass the three distinct types of distances probed by DESI BAO measurements, including $D_{\mathrm{H}}$ (left), $D_{\mathrm{M}}$ (middle), and $D_{\mathrm{V}}$ (right). The error bars represent $1\sigma$ uncertainties.
\textit{Lower panel:} Difference between the model prediction and data point for each BAO measurement, normalized by the observational uncertainties. The predictions for $\Lambda$CDM, $\widetilde{\Lambda}$CDM, and $e\widetilde{\Lambda}$CDM are shown by filled, cross-shaped markers, and plus signs, respectively.}
\end{figure*}

In the asymptotically safe quantum gravity scenario, the coupling parameters $G$ and $\Lambda$ evolve with redshift as
\begin{align}
\frac{G(z)}{G_0} &= (1+z)^{-\delta_G}, \\
\frac{\Lambda(z)}{\Lambda_0} &= (1+z)^{\delta_\Lambda},
\end{align}
where $\delta_G$ and $\delta_\Lambda$ are small critical indices ($\delta_G, \delta_\Lambda \ll 1$). Substituting these relations into the energy density expressions yields the effective forms
\begin{align}
\left(\frac{G(z)}{G_0}\right)\rho_{\mathrm{m}}(z) &= \rho_{\mathrm{m0}}(1+z)^{3-\delta_G}, \\
\left(\frac{G(z)}{G_0}\right)\rho_{\mathrm{r}}(z) &= \rho_{\mathrm{r0}}(1+z)^{4-\delta_G}, \\
\left(\frac{G(z)}{G_0}\right)\rho_{\Lambda}(z) &= \rho_{\Lambda0}(1+z)^{\delta_\Lambda}.
\end{align}
Here, $\delta_G$ is taken to be identical for both matter and radiation sectors, as their deviations originate solely from the same running of $G(z)$.

The generalized energy conservation equation in this framework reads
\begin{equation}
(1+z)\frac{\mathrm{d}E^2(z)}{\mathrm{d}z} =
3\Omega_{\mathrm{m0}}(1+z)^{3-\delta_G}
+ 4\Omega_{\mathrm{r0}}(1+z)^{4-\delta_G},
\end{equation}
where $E(z)\equiv H(z)/H_0$ and $\Omega_{\mathrm{m0}}$, $\Omega_{\mathrm{r0}}$ are the current fractional energy densities of matter and radiation, respectively.  
Combining this equation with the modified Friedmann relation (Eq.~\ref{eq:2.10} in the main text) yields the approximate low-redshift relation between the two critical indices
\begin{equation}
\delta_\Lambda \simeq \delta_G
\left(\frac{\Omega_{\mathrm{m0}}+\Omega_{\mathrm{r0}}}{\Omega_{\Lambda0}}\right)
\approx 0.47\,\delta_G.
\end{equation}
A non-zero $\delta_G$ or $\delta_\Lambda$ thus implies an energy exchange between the matter and dark energy sectors, while conserving the total energy budget of the universe.

The $\widetilde{\Lambda}$CDM model corresponds to the case where the two indices are related as $\delta_\Lambda \simeq 0.47\,\delta_G$, forming a single-parameter extension of the standard $\Lambda$CDM scenario.  
When $\delta_G$ and $\delta_\Lambda$ are treated as independent, the resulting two-parameter framework defines the extended $e\widetilde{\Lambda}$CDM model, which allows a more general coupling between the gravitational and dark energy sectors.

\section{Comparison of $\widetilde{\Lambda}$CDM and $e\widetilde{\Lambda}$CDM for DESI DR2}\label{appendixB}

To clarify the role of DESI data in the $\widetilde{\Lambda}$CDM and $e\widetilde{\Lambda}$CDM scenarios, we compare in Fig.~\ref{fig4} the theoretical distance predictions of the $\Lambda$CDM, $\widetilde{\Lambda}$CDM, and $e\widetilde{\Lambda}$CDM models with the observed cosmic distance measurements. Since DESI data alone provide relatively weak constraints, we focus on the joint analysis with CMB. Using CMB+DESI, we evaluate the best-fit predictions for three types of rescaled distances derived from DESI BAO data. These include the angle-averaged distance $D_{\mathrm{V}}$, the transverse comoving distance $D_{\mathrm{M}}$, and the Hubble horizon $D_{\mathrm{H}}$. We find that the distance predictions of $\Lambda$CDM and $\widetilde{\Lambda}$CDM remain in close agreement, which shows that the $\widetilde{\Lambda}$CDM framework is broadly consistent with $\Lambda$CDM. In contrast, the predictions of $\Lambda$CDM and $e\widetilde{\Lambda}$CDM reveal visible differences, implying possible interactions in the extended model. As shown in the lower panel of Fig.~\ref{fig4}, at most redshift points, the predictions of the $e\widetilde{\Lambda}$CDM model are within the $1\sigma$ range of the data. In particular, the two data points that depart from the $\Lambda$CDM and $\widetilde{\Lambda}$CDM predictions are $D_{\mathrm{M}}/(r_{\mathrm{d}}z^{2/3})$ at $z=0.71$ and $D_{\mathrm{H}}/(r_{\mathrm{d}}z^{-2/3})$ at $z=0.51$. Interestingly, while the value of $D_{\mathrm{H}}/(r_{\mathrm{d}}z^{-2/3})$ at $z=0.51$ also deviates from the $e\widetilde{\Lambda}$CDM prediction, the extended model provides a better description of $D_{\mathrm{M}}/(r_{\mathrm{d}}z^{2/3})$ at $z=0.71$ compared to $\Lambda$CDM. Overall, except for $D_{\mathrm{M}}/(r_{\mathrm{d}}z^{2/3})$ at $z=1.32$ and the measurements of $D_{\mathrm{M}}/(r_{\mathrm{d}}z^{2/3})$ and $D_{\mathrm{H}}/(r_{\mathrm{d}}z^{-2/3})$ at $z=0.51$, all other BAO observations are in good agreement with the best-fit predictions of the $e\widetilde{\Lambda}$CDM model.

\bibliography{main}
\end{document}